\documentclass[12pt,a4paper]{article}
\usepackage{a4wide,cite}

\newcommand{\Li}[2]{{\mbox{Li}}_{#1}\left(#2\right)}
\newcommand{\Cl}[2]{{\mbox{Cl}}_{#1}\left(#2\right)}
\newcommand{\Ls}[2]{{\mbox{Ls}}_{#1}\left(#2\right)}

\newcommand{\be}{\begin{equation}}
\newcommand{\ee}{\end{equation}}
\newcommand{\bea}{\begin{eqnarray}}
\newcommand{\eea}{\end{eqnarray}}
\newcommand{\ep}{\varepsilon}

\newcommand{\nn}{\nonumber}

\renewcommand{\thefootnote}{\fnsymbol{footnote}}

\begin{document}
%\pagestyle{myheadings}
%\markboth{\Draft{Version}}{\Draft{Version}}

 \thispagestyle{empty}
 \begin{flushright}
 {MZ-TH/99-44} \\[3mm]
 {hep-ph/9910224} \\[3mm]
 {October 1999}
 \end{flushright}
 \vspace*{2.0cm}
 \begin{center}
 {\Large \bf
Explicit results for all orders of the $\ep$-expansion\\[2mm]
of certain massive and massless diagrams}
 \end{center}
 \vspace{1cm}
 \begin{center}
 A.I.~Davydychev\footnote{Alexander von Humboldt fellow. On leave from
                 Institute for Nuclear Physics, Moscow State University,
                 119899 Moscow, Russia.
Email: davyd@thep.physik.uni-mainz.de}
\\
 \vspace{1cm}
{\em
 Department of Physics,
 University of Mainz, \\
 Staudingerweg 7,
 D-55099 Mainz, Germany}
\end{center}
 \hspace{3in}
\begin{abstract}
An arbitrary term of the
$\ep$-expansion of dimensionally regulated off-shell massless one-loop 
three-point Feynman diagram is expressed in terms of log-sine integrals
related to the polylogarithms. Using magic connection between these 
diagrams and two-loop massive vacuum diagrams, the $\ep$-expansion
of the latter is also obtained, for arbitrary values of the masses.
The problem of analytic continuation is also discussed.
\end{abstract}

%%%%%%%%%%%%%%%%%%%%%%%%%%%%%%%%%%%%%%%%%%%%%%%%%%%%%%%%%%%%%%%%%%%%%%%
\newpage
\renewcommand{\thefootnote}{\arabic{footnote}}
\setcounter{footnote}{0}

{\bf 1.}
In this paper we shall discuss some issues related to the evaluation 
of Feynman integrals in the framework of dimensional 
regularization \cite{dimreg},
when the space-time dimension is $n=4-2\ep$.
Sometimes it is possible to present results valid for an
arbitrary $\ep$, usually in terms of hypergeometric functions.  
However, for practical purposes the coefficients
of the expansion in $\ep$ are important.
In particular, in multi-loop calculations higher terms
of the $\ep$-expansion of one- and two-loop functions are needed, since
one can get contributions where these functions are
multiplied by singular factors containing poles in $\ep$.
Such poles may appear not only due to factorizable loops,
but also as a result of applying the integration by parts \cite{ibp}
or other techniques \cite{Tarasov}.

For the one-loop two-point function $J^{(2)}(n;\nu_1,\nu_2)$ with 
external momentum $k_{12}$, masses $m_1$ and $m_2$ 
and unit powers of propagators $\nu_1=\nu_2=1$,
we have obtained the following result for an arbitrary term
of the $\ep$-expansion 
%eq.~(21) of 
(see ref.~\cite{Crete}):
\bea
\label{2pt_res2}
J^{(2)}(4\!-\!2\ep;1,1) = \mbox{i}\pi^{2-\ep}
\frac{\Gamma(1\!+\!\ep)}{2(1-2\ep)}
\left\{ \frac{m_1^{-2\ep} \!+\! m_2^{-2\ep}}{\ep}  
+ \frac{m_1^2\!-\!m_2^2}{\ep \; k_{12}^2}
\left( m_1^{-2\ep} \!-\! m_2^{-2\ep} \right)
\right.
\hspace{20mm}
\nn \\
\left.
%+ \frac{2^{-2\ep}(m_1 m_2 \sin\tau_{12})^{1-2\ep}}{(k_{12}^2)^{1-\ep}}
- \left[\Delta(m_1^2,m_2^2,k_{12}^2)\right]^{1/2-\ep}
\sum\limits_{j=0}^{\infty} \frac{(2\ep)^j}{j!}
%\nn \\
%&& \left. \times
\left[ \mbox{Ls}_{j+1}(2\tau'_{01})\!
+\! \mbox{Ls}_{j+1}(2\tau'_{02})\!
- \!2\mbox{Ls}_{j+1}(\pi)
\right] \!\frac{}{}\! \right\} ,   
\hspace*{3mm}
\eea
where
\be
\label{two-point}
\cos\tau_{12} = \frac{m_1^2+m_2^2-k_{12}^2}{2m_1 m_2}, \hspace{5mm}
\cos\tau'_{01} = \frac{m_1^2-m_2^2+k_{12}^2}{2m_1 \sqrt{k_{12}^2}}, \hspace{5mm}
\cos\tau'_{02} = \frac{m_2^2-m_1^2+k_{12}^2}{2m_2 \sqrt{k_{12}^2}} .
\ee
The angles $\tau'_{0i}$ ($i=1,2$) are related to the angles $\tau_{0i}$ 
used in refs.~\cite{DD,Crete} via $\tau'_{0i}=\pi/2-\tau_{0i}$.
The ``triangle'' function $\Delta$ is defined as
\be
\label{Delta}
\Delta(x,y,z)=2xy+2yz+2zx-x^2-y^2-z^2=-\lambda(x,y,z),
\ee
where $\lambda(x,y,z)$ is the K\"allen function.
The coefficient of $\ep^j$ has a closed form in terms
of log-sine integrals (see in \cite{Lewin}, chapter~7.9),
\be
\label{Ls_j}
\mbox{Ls}_j(\theta)\equiv - \int\limits_0^{\theta}
\mbox{d}\theta' \ln^{j-1}\left|2\sin\frac{\theta'}{2}\right| .
\ee
In particular, $\Ls{1}{\theta}=-\theta$ and $\Ls{2}{\theta}=\Cl{2}{\theta}$,
where 
\be
\label{Cl_j}
\Cl{j}{\theta}=\left\{ 
\begin{array}{l} 
\mbox{Im}\left[\Li{j}{e^{{\rm i}\theta}}\right] 
=\left[ \Li{j}{e^{{\rm i}\theta}} - \Li{j}{e^{-{\rm i}\theta}} \right]/(2\mbox{i}) , 
\hspace{3mm} j \;\; \mbox{even} \\[2mm]
\mbox{Re}\left[\Li{j}{e^{{\rm i}\theta}}\right]
=\left[ \Li{j}{e^{{\rm i}\theta}} + \Li{j}{e^{-{\rm i}\theta}} \right]/2 ,
\hspace{6mm} j \;\; \mbox{odd} 
\end{array}
\right.
\ee
is the Clausen function (see in \cite{Lewin}), $\mbox{Li}_j$ is the polylogarithm.
Note that the values of $\Ls{j}{\pi}$ can be expressed in terms of 
Riemann's $\zeta$ function, see eqs.~(7.112)--(7.113) of \cite{Lewin}.
Moreover, the infinite sum with $\Ls{j}{\pi}$ in (\ref{two-point})
can be converted into $\Gamma$ functions (see eq.~(\ref{phi=pi/2}) below).

The arguments of Ls functions in eq.~(\ref{two-point}) have simple 
geometrical interpretation \cite{DD}.
We note that $\tau_{12}+\tau'_{01}+\tau'_{02}=\pi$
(equivalently, $\tau_{01}+\tau_{02} = \tau_{12}$).
Therefore, $\tau_{12}$, $\tau'_{01}$ and $\tau'_{02}$ can be 
understood as the angles 
of a triangle whose sides are $m_1$, $m_2$ and $\sqrt{k_{12}^2}$,
whereas the area of this triangle is 
$\textstyle{1\over4}\sqrt{\Delta(m_1^2,m_2^2,k_{12}^2)}$.
The $\ep$-expansion (\ref{two-point}) is directly applicable in the
region where $\Delta(m_1^2,m_2^2,k_{12}^2)\geq 0$, i.e.\ when
$(m_1-m_2)^2\leq k_{12}^2\leq(m_1+m_2)^2$. In other regions,
the proper analytic continuation should be constructed.
We note that the result for the $\ep$-term
was obtained in \cite{NMB}.
For the case $m_1=0$ ($m_2\equiv m$), the
first terms of the expansion (up to $\ep^3$) are presented
in eq.~(A.3) of \cite{FJTV}.

We shall show that similar explicit results
can be constructed for the off-shell massless one-loop three-point function
with external momenta $p_1$, $p_2$ and $p_3$ ($p_1+p_2+p_3=0$),
\be
\label{defJ}
J (n; \; \nu_1  ,\nu_2  ,\nu_3 | p_1^2, p_2^2, p_3^2) \equiv \int
 \frac{\mbox{d}^n r}{ \left[(p_2 -r )^2\right]^{\nu_1}  
\left[(p_1 +r )^2\right]^{\nu_2}
      (r^2)^{\nu_3} } ,
\ee
as well as for the two-loop vacuum diagram with arbitrary masses
$m_1$, $m_2$ and $m_3$,
\be
\label{defI}
I(n; \; \nu_1, \nu_2, \nu_3 | m_1^2, m_2^2, m_3^2)
\equiv  \int \int \frac{\mbox{d}^n p \;\; \mbox{d}^n q}
      {\left( p^2 - m_1^2 \right)^{\nu_1}
       \left( q^2 - m_2^2 \right)^{\nu_2}
       \left[ (p-q)^2 - m_3^2 \right]^{\nu_3}} .
\ee
According to the magic connection \cite{DT2}, they are closely 
related to each other.
To be specific, we shall consider the case $\nu_1=\nu_2=\nu_3=1$, 
although the approach can be also applied 
to arbitrary integer values of $\nu_i$.

\vspace{5mm}

{\bf 2.}
Thus, our purpose is to obtain an arbitrary term of the $\ep$-expansion 
of dimensionally regulated integrals
$J(4-2\ep;1,1,1)$ and $I(4-2\ep;1,1,1)$ defined in eqs.~(\ref{defJ})
and (\ref{defI}), respectively.
As in \cite{DT2}, we assume that $p_i^2\leftrightarrow m_i^2$.
Moreover, below we shall omit the arguments $p_i^2$  
and $m_i^2$ in the integrals $J$ and $I$, respectively.
We shall mainly be interested in the region where
all $p_i^2$ are positive (time-like), whereas 
\be
\Delta(p_1^2,p_2^2,p_3^2)\equiv -\lambda(p_1^2,p_2^2,p_3^2)\geq 0.
\ee

According to the magic connection 
(we use eq.~(16) of \cite{DT2}, with changed sign of $\ep$),
\be
J(4-2\ep;1,1,1)=\pi^{-3\ep}\; \mbox{i}^{1+2\ep}\; 
\left( p_1^2 p_2^2 p_3^2 \right)^{-\ep}\;
\frac{\Gamma(1+\ep)}{\Gamma(1-2\ep)}\;
I(2+2\ep;1,1,1) .
\ee
Then, for $I(2+2\ep;1,1,1)$ we can use an exact result (4.12)--(4.13)
from \cite{DT1} (see also in \cite{FJJ}), 
in terms of hypergeometric $_2F_1$ functions.
Using the well-known transformation
\be
\left. _2F_1\left( \begin{array}{c} \ep, 1 \\ 1/2+\ep \end{array}
\right| z \right)
= \frac{1}{1-z}\; 
\left. _2F_1\left( \begin{array}{c} 1, 1/2 \\ 1/2+\ep \end{array}
\right| \frac{z}{z-1} \right) ,
\ee
and then changing
$\ep\to 1-\ep$, we arrive at
\bea
\label{J-2F1}
J(4-2\ep;1,1,1) &=& 2 \pi^{2-\ep}\; \mbox{i}^{1+2\ep}\;
\left( p_1^2 p_2^2 p_3^2 \right)^{-\ep}\;
\frac{\Gamma^2(1-\ep)\;\Gamma(\ep)}{\Gamma(2-2\ep)}\;
\nn \\
&& \times\left\{
\frac{(p_1^2 p_2^2)^{\ep}}{p_1^2+p_2^2-p_3^2}\;
\left. _2F_1\left( \begin{array}{c} 1, 1/2 \\ 3/2-\ep \end{array}
\right| -\frac{\Delta(p_1^2,p_2^2,p_3^2)}{(p_1^2+p_2^2-p_3^2)^2} \right)
\right.
\nn \\
&& \hspace*{4mm} 
+\frac{(p_2^2 p_3^2)^{\ep}}{p_2^2+p_3^2-p_1^2}\;
\left. _2F_1\left( \begin{array}{c} 1, 1/2 \\ 3/2-\ep \end{array}
\right| -\frac{\Delta(p_1^2,p_2^2,p_3^2)}{(p_2^2+p_3^2-p_1^2)^2} \right)
\nn \\
&&  \hspace*{4mm}
+\frac{(p_3^2 p_1^2)^{\ep}}{p_3^2+p_1^2-p_2^2}\;
\left. _2F_1\left( \begin{array}{c} 1, 1/2 \\ 3/2-\ep \end{array}
\right| -\frac{\Delta(p_1^2,p_2^2,p_3^2)}{(p_3^2+p_1^2-p_2^2)^2} \right)
\nn \\
&& \hspace*{3mm}
\left. \frac{}{}
-\pi \frac{\Gamma(2-2\ep)}{\Gamma^2(1-\ep)}\;
\left[ \Delta(p_1^2,p_2^2,p_3^2)\right]^{-1/2+\ep}\;
\Theta_{123} \right\} \; ,
\eea
with
\be
\label{Theta123}
\Theta_{123} \equiv
\theta(p_1^2+p_2^2-p_3^2)\; \theta(p_2^2+p_3^2-p_1^2)\;
\theta(p_3^2+p_1^2-p_2^2) 
\ee
(cf.\ eq.~(4.13) of \cite{DT1}).

Of course, the same result (\ref{J-2F1}) can be also obtained by putting
$\nu_1=\nu_2=\nu_3=1$ in general results from \cite{triangle,JPA} and
using reduction formulae for the occurring $F_4$ functions,
basically repeating the steps done in \cite{DT1} for 
$I(4-2\ep;1,1,1)$. 

It is convenient to introduce the angles $\phi_i$ ($i=1,2,3$) such that
\be
\cos\phi_1=\frac{p_2^2+p_3^2-p_1^2}{2\sqrt{p_2^2 p_3^2}},
\hspace{4mm}
\cos\phi_2=\frac{p_3^2+p_1^2-p_2^2}{2\sqrt{p_3^2 p_1^2}},
\hspace{4mm}
\cos\phi_3=\frac{p_1^2+p_2^2-p_3^2}{2\sqrt{p_1^2 p_2^2}},
\ee
so that $\phi_1+\phi_2+\phi_3=\pi$,
and the arguments of $_2F_1$ functions in eq.~(\ref{J-2F1}) are
nothing but minus $\tan^2\phi_i$. 
Note that the angles $\theta_i$ from \cite{DT1,DT2} are related
to $\phi_i$ as $\theta_i=2\phi_i$.
By analogy with the two-point case (\ref{two-point}), the angles $\phi_i$ 
can be understood as the angles of a triangle 
whose sides are $\sqrt{p_1^2}$, $\sqrt{p_2^2}$ and $\sqrt{p_3^2}$,
whereas its area is $\textstyle{1\over4}\sqrt{\Delta(p_1^2,p_2^2,p_3^2)}$.

\vspace{5mm}

{\bf 3.}
The crucial step in constructing the $\ep$-expansion is the formula
\bea
\label{sum_Ls}
\sum\limits_{j=0}^{\infty}\frac{(-2\ep)^j}{j!}\; \Ls{j+1}{2\phi}
&=& -\frac{2^{1-2\ep}\tan\phi}{(1-2\ep)\sin^{2\ep}\phi}\;
\left. _2F_1\left( \begin{array}{c} 1, 1/2 \\ 3/2-\ep \end{array}
\right| -\tan^2\phi \right)
\nn \\
&& -2\pi\; \frac{\Gamma(1-2\ep)}{\Gamma^2(1-\ep)}\;
\theta(-\cos\phi) \; .
\eea
Using the definition (\ref{Ls_j}),
we see that the l.h.s. of eq.~(\ref{sum_Ls}) is nothing but
\be
\sum\limits_{j=0}^{\infty}\frac{(-2\ep)^j}{j!}\; \Ls{j+1}{2\phi}
=-2^{1-2\ep} \int\limits_0^{\phi}\mbox{d}\tau\; \sin^{-2\ep}\tau \; .
\ee
Evaluating this integral (treating the cases $0\leq\phi<\pi/2$
and $\pi/2<\phi<\pi$ separately), we arrive at the r.h.s. of 
eq.~(\ref{sum_Ls}), q.e.d.

In the special case $\phi=\pi/2$ we get (see in \cite{Lewin,Crete})
\be
\label{phi=pi/2}
\sum\limits_{j=0}^{\infty} \frac{(-2\ep)^j}{j!}\;
\mbox{Ls}_{j+1}(\pi) = -\pi\; \frac{\Gamma(1-2\ep)}{\Gamma^2(1-\ep)} \; .
\ee

Now, using eq.~(\ref{sum_Ls}) for all three $_2F_1$ functions 
occurring in eq.~(\ref{J-2F1}),
identifying $\Theta_{123}$ (see eq.~(\ref{Theta123})) as
\be
\Theta_{123}=
1-\theta(-\cos\phi_1)-\theta(-\cos\phi_2)
-\theta(-\cos\phi_3),
\ee
then using eq.~(\ref{phi=pi/2}) and shifting $j\to j+1$, we arrive at
the following $\ep$-expansion:
\bea
\label{conjecture}
J(4-2\ep;1,1,1) &=& 2 \pi^{2-\ep} \; \mbox{i}^{1+2\ep}\;
\frac{\Gamma(1+\ep)\Gamma^2(1-\ep)}{\Gamma(1-2\ep)}\;
\frac{\left[ \Delta(p_1^2,p_2^2,p_3^2)\right]^{-1/2+\ep}}
     {(p_1^2 p_2^2 p_3^2)^{\ep}}
\nn \\
&& \!\times \sum\limits_{j=0}^{\infty} \frac{(-2\ep)^j}{(j\!+\!1)!}
\left[ \Ls{j+2}{2\phi_1} \!+\! 
\Ls{j+2}{2\phi_2} \!+\! \Ls{j+2}{2\phi_3}
\!-\! 2 \Ls{j+2}{\pi} \right] .
%\nn \\
\hspace{6mm}
\eea
The shift of $j$ was possible, since
\be
\Ls{1}{2\phi_1} \!+\!
\Ls{1}{2\phi_2} \!+\! \Ls{1}{2\phi_3}
\!-\! 2 \Ls{1}{\pi}
= - 2 (\phi_1+\phi_2+\phi_3-\pi) = 0 .
\ee 

The $\ep$-expansion of the two-loop vacuum integral
$I(4-2\ep;1,1,1)$ can be obtained via the magic connection, i.e.\ 
by substituting eq.~(\ref{conjecture}) into eq.~(23) of \cite{DT2}:
\bea
\label{conjecture2}
I(4-2\ep;1,1,1)&=& \pi^{4-2\ep}\; \frac{\Gamma^2(1+\ep)}{(1-\ep)(1-2\ep)}\;
\left\{ \left[\Delta(m_1^2,m_2^2,m_3^2)\right]^{1/2-\ep} \frac{}{}
\right.
\nn \\
&& \times 
\sum\limits_{j=0}^{\infty} \frac{(2\ep)^j}{(j\!+\!1)!}
\left[ \Ls{j+2}{2\phi_1} \!+\!
\Ls{j+2}{2\phi_2} \!+\! \Ls{j+2}{2\phi_3}
\!-\! 2 \Ls{j+2}{\pi} \right] 
\nn \\
&& \left.
-\frac{1}{2\ep^2}
\left[ \frac{m_1^2+m_2^2-m_3^2}{(m_1^2 m_2^2)^{\ep}}
+\frac{m_2^2+m_3^2-m_1^2}{(m_2^2 m_3^2)^{\ep}}
+\frac{m_3^2+m_1^2-m_2^2}{(m_3^2 m_1^2)^{\ep}}
\right] \right\} \; .
\eea

When the masses are equal, $m_1=m_2=m_3\equiv m$
(this also applies to the symmetric case $p_1^2=p_2^2=p_3^2\equiv p^2$), 
the three angles $\phi_i$ are all equal to $\pi/3$,
whereas $\Delta(m^2,m^2,m^2)=3m^4$.
Therefore, in this case the r.h.s. of eq.~(\ref{conjecture2}) becomes
\be
\label{equal_m}
\pi^{4-2\ep}\; \frac{\Gamma^2(1+\ep)\; m^{2-4\ep}}{(1-\ep)(1-2\ep)}
\left\{ 3^{1/2-\ep} \sum\limits_{j=0}^{\infty}
\frac{(2\ep)^j}{(j\!+\!1)!}
\left[ 3\Ls{j+2}{\frac{2\pi}{3}}- 2\Ls{j+2}{\pi} \right] 
-\frac{3}{2\ep^2} \right\} .
\ee
For instance, in the contribution of order $\ep$ the transcendental 
constant $\Ls{3}{2\pi/3}$ appears. This constant (and its connection
with the inverse tangent integral value $\mbox{Ti}_3(1/\sqrt{3})$) was 
discussed in detail in \cite{DT2}.
The fact that $\Ls{3}{2\pi/3}$ occurs in certain two-loop
on-shell integrals has been noticed in \cite{FKK}.
Moreover, in \cite{FKnew} it was observed that the higher-$j$
terms from (\ref{equal_m}) form a basis for certain on-shell
integrals with a single mass parameter.

Note that the structure of eq.~(\ref{conjecture})
is very similar to that
of the two-point function with masses (\ref{two-point}).
For the two lowest orders ($\ep^0$ and $\ep^1$), we reproduce
eqs.~(9)--(10) from \cite{DT2}.
Useful representations for the $\ep^0$ terms of both types of diagrams 
can also be found in \cite{ep0}.
We note that the $\ep$-term of the one-loop massive three-point
function was calculated in \cite{NMB}, whereas the massless case
was considered in \cite{UD3}. 

Moreover, in eq.~(26)
of \cite{UD3} a one-fold integral representation 
for $J(4-2\ep;1,1,1)$ is presented
(for its generalization, see eq.~(7) of \cite{DT2}).
Expanding the integrand in $\ep$, we were able to confirm the
$\ep$-expansion (\ref{conjecture}) numerically. 

\vspace{5mm}

{\bf 4.}
We have shown that the compact structure
of the coefficients of the $\ep$-expansion
of the two-point function (\ref{two-point}), 
in terms of log-sine integrals,
also takes place for
the massless off-shell three-point function (\ref{conjecture})
and two-loop massive vacuum diagrams (\ref{conjecture2}).
It is likely that a further generalization of these results 
is possible, e.g.\ for the three-point function with different masses
and some two-point integrals with two (and more) loops. In particular, 
numerical analysis of the coefficients of the expansion of
certain two-point on-shell integrals and three-loop vacuum integrals
\cite{FKnew} shows that
in some cases the values of generalized log-sine integrals 
$\mbox{Ls}_j^{(l)}$ (see eq.~(7.14) of \cite{Lewin}) may be involved.

The fact that the generalization of $\mbox{Ls}_2=\mbox{Cl}_2$
goes in the $\mbox{Ls}_j$ direction, rather than in $\mbox{Cl}_j$
direction (see eq.~(\ref{Cl_j})),
is very interesting. There is another example \cite{UD2} (see also
in \cite{Bro93}), 
the off-shell massless ladder three- and four-point diagrams with
an arbitrary number of loops, when such a generalization went
in the $\mbox{Cl}_j$ direction. Just as an illustration, we can
present the result for the $L$-loop function $\Phi^{(L)}(x,y)$
(for the definition, see eqs.~(12) and (21) of \cite{UD2};
$x\leftrightarrow p_1^2/p_3^2$, $y\leftrightarrow p_2^2/p_3^2$) 
in the case $y=x$,
which is valid when $\Delta(p_1^2,p_2^2,p_3^2)<0$:
\be
\Phi^{(L)}(x,x)=\frac{(2L)!}{(L!)^2}\; \frac{1}{x \sin\theta}\;
\Cl{2L}{\theta} , 
\hspace{6mm}
\theta=\arccos\left(1-\frac{1}{2x}\right) .
\ee
When $x=1$ ($p_i^2=p^2$) this yields the $\Cl{2L}{\pi/3}/\sqrt{3}$ 
structures. 
It could be also noted that the the two-loop non-planar (crossed) 
three-point diagram gives in this case the square of 
the one-loop function, $(\Cl{2}{\theta})^2$ (cf.\ eq.~(23) of \cite{UD3}),
leading to the structute $(\Cl{2}{\pi/3})^2$ in the symmetric
($p_i^2=p^2$) case. Recently, these constants have been also found
in massive three-loop calculations \cite{rho,Bro99} (see also in
\cite{GKP}).

The representations (\ref{two-point}), (\ref{conjecture}) 
and (\ref{conjecture2}) are 
directly applicable to the case when the triangle function 
$\Delta$ given by eq.~(\ref{Delta}) is positive. When $\Delta$ is negative,
we need to construct proper analytic continuation of $\mbox{Ls}_j$
functions. For $j=2$ this is simple, since
$\Ls{2}{\theta}=\Cl{2}{\theta}$ and we can use the definition (\ref{Cl_j}).
Similarly one can deal with higher $\mbox{Cl}_j$ functions.
Let us consider the situation with analytic continuation 
of higher $\mbox{Ls}_j$ functions.
For $j=3$, $\Ls{3}{\theta}$ can be expressed in terms of
the imaginary part of $\Li{3}{1-e^{{\rm i}\theta}}$, see 
in \cite{Lewin}\footnote{A factor ${\textstyle{1\over2}}$
is missing in front of $\Ls{3}{\theta}$ in
eq.~(49) in p.~298 of \cite{Lewin}, cf.\ eq.~(6.56).}.
Using this fact, we can re-construct eq.~(16) of the preprint version 
of \cite{DT2}, which gives the analytic continuation
of the $\ep$-term of eqs.~(\ref{conjecture}) and (\ref{conjecture2}).
Then, the imaginary part of $\Li{4}{1-e^{{\rm i}\theta}}$ 
is already a mixture of $\Ls{4}{\theta}$ and $\Cl{4}{\theta}$,
see in \cite{Lewin}\footnote{In eq.~(7.67) of \cite{Lewin}, 
as well as in eq.~(36) in p.~301, the coefficient of 
$\log^2\left(2\sin{\textstyle{1\over2}}\theta\right)\Cl{2}{\theta}$
should be $-\textstyle{1\over2}$ (rather than $+\textstyle{3\over2}$).},
whereas its real part involves the generalized log-sine integral
$\mbox{Ls}_4^{(1)}(\theta)$ (see eq.~(35) in p.~301 of \cite{Lewin}).
Its value at $\theta=2\pi/3$
also occurs in on-shell integrals considered in \cite{FKnew}
as is shown to be connected with $V_{3,1}$ from \cite{Bro99}.
The construction of analytic continuation of higher $\mbox{Ls}_j$ 
functions
is more cumbersome. In fact, it may require including the generalized
polylogarithms (see, e.g., in ref.~\cite{RV}).

\vspace{3mm}

$\hspace*{-2mm}$
{\bf Acknowledgements.} I am grateful to M.Yu.~Kalmykov for many useful
discussions and informing me about the results of \cite{FKnew} before
publication. I wish to thank J.B.~Tausk for useful comments.
The research was supported by the Alexander von Humboldt Foundation.
Partial support from RFBR grant No.~98--02--16981
is acknowledged.

%%%%%%%%%%%%%%%%%%%%%%%%%%%%%%%%%%%%%%%%%%%%%%%%%%%%%%%%%%%%%%%%%%%%%%%


\begin{thebibliography}{99}

\bibitem{dimreg}
G.~'tHooft and M.~Veltman,
{\em Nucl.~Phys.} {\bf B44} (1972) 189;\\
C.G.~Bollini and J.J.~Giambiagi,
{\em Nuovo~Cimento} {\bf 12B} (1972) 20; \\
J.F.~Ashmore,
{\em Lett.~Nuovo~Cim.} {\bf 4} (1972) 289;\\
G.M.~Cicuta and E.~Montaldi,
{\em Lett.~Nuovo~Cim.} {\bf 4} (1972) 329.

\bibitem{ibp}
F.V.~Tkachov, {\em Phys.~Lett.} {\bf 100B} (1981) 65; \\   
K.G.~Chetyrkin and F.V.~Tkachov,
   {\em Nucl.~Phys.} {\bf B192} (1981) 159.

\bibitem{Tarasov}
O.V.~Tarasov, {\em Nucl.~Phys.} {\bf B502} (1997) 455.

\bibitem{Crete}
A.I. Davydychev, Mainz preprint MZ-TH/99-30 (hep-th/9908032).

\bibitem{DD}
A.I.~Davydychev and R.~Delbourgo,
{\em J.~Math.~Phys.} {\bf 39} (1998) 4299.

\bibitem{Lewin}
L.~Lewin, {\em Polylogarithms and associated functions},
      North Holland, 1981.

\bibitem{NMB}
U.~Nierste, D.~M\"uller and M. B\"ohm, {\em Z.~Phys.}
      {\bf C57} (1993) 605.

\bibitem{FJTV}
J.~Fleischer, F.~Jegerlehner, O.V.~Tarasov and O.L. Veretin,
{\em Nucl.~Phys.} {\bf B539} (1999) 671.

\bibitem{DT2}
A.I.~Davydychev and J.B.~Tausk,
{\em Phys.~Rev.} {\bf D53} (1996) 7381.

\bibitem{DT1}
A.I.~Davydychev and J.B.~Tausk,
{\em Nucl.~Phys.} {\bf B397} (1993) 123.

\bibitem{FJJ}
C.~Ford, I.~Jack and D.R.T.~Jones,
{\em Nucl.~Phys.} {\bf B387} (1992) 373.

\bibitem{triangle}
E.E.~Boos and A.I. Davydychev,
{\em Vestn. Mosk. Univ.} (Ser.3) {\bf 28}, No.3 (1987) 8. 
%[{\em Mosc. Univ. Phys. Bull. {\bf 42}, No.3 (1987) 6]

\bibitem{JPA}
A.I. Davydychev, {\em J. Phys.} {\bf A25} (1992) 5587.

\bibitem{FKK}
J.~Fleischer, M.Yu.~Kalmykov and A.V.~Kotikov,
{\em Phys.~Lett.} {\bf B462} (1999) 169; \\
J.~Fleischer and M.Yu.~Kalmykov, hep-ph/9907431.

\bibitem{FKnew}
J.~Fleischer and M.Yu.~Kalmykov, Preprint BI-TP 99/34
(hep-ph/9910223).

\bibitem{ep0}
J.S.~Ball and T.-W.~Chiu, {\em Phys.~Rev.} {\bf D22} (1980) 2550;
        {\bf D23} (1981) 3085(E); \\
J.J.~van der Bij and M.~Veltman, {\em Nucl.~Phys.} {\bf B231} (1984) 205; \\
H.J.~Lu and C.A.~Perez, preprint SLAC-PUB-5809, 1992; \\
N.I.~Ussyukina and A.I.~Davydychev, {\em Phys.~Lett.} {\bf B298} (1993) 363.

\bibitem{UD3}
N.I.~Ussyukina and A.I.~Davydychev,
{\em Phys.~Lett.} {\bf B332} (1994) 159.

\bibitem{UD2} N.I.~Ussyukina and A.I.~Davydychev,
{\em Phys.~Lett.} {\bf B305} (1993) 136.

\bibitem{Bro93} D.J.~Broadhurst,
{\em Phys.~Lett.} {\bf B307} (1993) 132.

\bibitem{rho} L.~Avdeev, J.~Fleischer, S.~Mikhailov and O.~Tarasov,
{\em Phys.~Lett.} {\bf B336} (1994) 560; {\bf B349} (1995) 597(E); \\
K.G.~Chetyrkin, J.H.~K\"uhn, M.~Steinhauser,
{\em Phys.~Lett.} {\bf B351} (1995) 331.

\bibitem{Bro99} D.J.~Broadhurst,
{\em Eur.~Phys.~J.} {\bf C8} (1999) 311.

\bibitem{GKP}
S.~Groote, J.G.~K\"orner, A.A.~Pivovarov,
{\em Phys.~Rev.} {\bf D60} (1999) 061701.

\bibitem{RV}
E.~Remiddi and J.A.M.~Vermaseren, perprint NIKHEF-99-005 
(hep-ph/9905237).

\end{thebibliography}
\end{document}